\documentclass{aa}

\usepackage{natbib}
\usepackage{graphicx}
\usepackage{txfonts}
\usepackage{subfigure}

\bibpunct{(}{)}{;}{a}{}{,}
\begin{document}

\title{\object{I Zw 1}: Decomposition of the Nearby QSO Host\thanks{Based on observations collected at the Very Large Telescope (UT1) of the 
European Southern Observatory, Paranal, Chile, under service mode project 67.B-0009.}}

\author{J. Scharw\"achter\inst{1} \and
           A. Eckart\inst{1} \and
           S. Pfalzner\inst{1} \and
           J. Moultaka\inst{1} \and
           C. Straubmeier\inst{1} \and
	   J. G. Staguhn\inst{2}
}

\offprints{J. Scharw\"achter, \email{scharw@ph1.uni-koeln.de}}

\institute{I. Physikalisches Institut, Universit\"at zu K\"oln, Z\"ulpicher Str. 77, 50937 K\"oln \and
NASA/Goddard Space Flight Center, Building 21, Code 685, Greenbelt, MD 20771}

\date{Received (date); accepted (date)}

\abstract{The paper presents an analysis of the nearby QSO host 
\object{I Zw 1} 
based on new J-band 
imaging data, obtained 
with ISAAC at the Very Large Telescope (VLT) of the European
Southern Observatory (ESO). 
As one of the best-studied QSOs with 
exceptional properties lying between those of QSOs, narrow-line
Seyfert~1s, and ultraluminous infrared galaxies,  
\object{I Zw 1} is a 
prime candidate for a detailed case study of a QSO host. 
With its high angular resolution and sensitivity, the new J-band image 
provides strong evidence for an interaction between \object{I Zw 1} and
the western companion galaxy. 
We present a procedure for subtracting the QSO nucleus from the
\object{I Zw 1} image to obtain an undisturbed view on the underlying host 
galaxy. Based on the derived surface brightness profile and the gas 
rotation curve, the host is classified as a 
high-surface-brightness Freeman type I galaxy with a central disk surface 
brightness significantly larger than that of normal spirals. 
The surface-brightness profile
is decomposed into a Plummer bulge and a Kuzmin disk with similar luminosities
and a bulge-to-disk scale length ratio of 0.29, comparable to that of
nearby spiral hosts with active galactic 
nuclei (AGN). Different models for the 
decomposition of the gas rotation curve are discussed. The 
resulting J-band mass-to-light ratio 
(M/L) of $(0.7\pm 0.4)\ M_{\sun}/L_{\sun}$ for the disk component
suggests a mean solar-type stellar population with a tendency towards young
stars.
The bulge exhibits a lower M/L of $(0.4\pm 0.2)\ M_{\sun}/L_{\sun}$, 
which
supports previous findings of enhanced starburst 
activity  
in the nuclear and circumnuclear region.
}

\maketitle

%%%%%%%%%%%%%%%%%%%%%%%%%%%%%%%%%%%%%%%%%%%%%%%%%%%%%%%%%%%%%%%%%%%%%

\section{Introduction\label{sec:intro}}

Due to their extreme properties in many physical respects,
active galactic nuclei have attracted much interest for a long time.  
Thanks to improvements in sensitivity and angular resolution,
the host galaxies have become a new focal point
\citep[e.g.][]{1984AJ.....89.1293B, 1995ApJ...439..588L, 1995PASP..107...91M, 1999ApJ...510..637H, 2000A&A...360..431M},
as host disks are considered as the fuel reservoirs
for their embedded active nuclei 
\citep[e.g.][]{1990Natur.345..679S, 2001AJ....121.1893E, 2002ApJ...569..624P}.
Structures and stellar populations
of the hosts seem to be closely entangled with 
the evolution and types of their active nuclei \citep{2001ApJ...559..147S}.
In order to minimize the effects of dust extinction, stellar properties of
host galaxies are preferentially observed in the near-infrared (NIR).
Nevertheless, the measured nuclear and 
circumnuclear fluxes are still
dominated by the high energy output of the embedded 
active nucleus. As an unresolved source, the active 
nucleus is usually removed by subtracting an appropriately
scaled point-spread function (PSF).
Most statistical studies of AGN hosts, however, 
lack an exact knowledge of
the AGN flux contribution so that the scaling is often  
rather arbitrary. It is determined either by assuming the residual 
central stellar component of the host to be flat or by applying 
some fitting algorithm 
\citep[e.g.][]{1995ApJ...450..486B, 1999ApJ...510..637H, 2001MNRAS.322..843P}. 
In contrast to these statistical approaches, this paper focusses on a 
detailed study of one nearby representative of QSO hosts,
the QSO host \object{I Zw 1}, for which some basic data are listed in 
Table~\ref{tab:zwdata}. 

\begin{table}
\caption{Basic data for \object{I Zw 1}.}
\label{tab:zwdata}
 $$
\begin{array}{lll}
	\hline
	\noalign{\smallskip}
	\mathrm{RA}_{2000} & \mathrm{00h53m34.9s} 
	& $\citet{1999PASP..111..438F}$ \\
	\mathrm{Dec}_{2000} & +12d41\arcmin 36\arcsec 
	& $\citet{1999PASP..111..438F}$\\
	\mathrm{Position\ Angle} & 135\degr  
	& $\citet{1998ApJ...500..147S}$\\
	\mathrm{Inclination} & (38\pm 5)\degr  
	& $\citet{1998ApJ...500..147S}$\\
	\mathrm{Redshift} & 0.0611 & $\citet{1997ApJ...478..144S}$\\
	\mathrm{Distance} & 242\ \mathrm{Mpc}^{\mathrm{a}} & \\
	\mathrm{Length\ Conversion\ Factor} 
	& 1 \arcsec \approx 1.2\ \mathrm{kpc} & \\
	\noalign{\smallskip}
	\hline
\end{array}
 $$
\begin{list}{}{}
	\item[$^{\mathrm{a}}$] $H_0=75$~km~s$^{-1}$~Mpc$^{-1}$ and 
	$q_0=0$ will be adopted throughout the paper.
\end{list}
\end{table}

This case study benefits from the fact that
a reliable subtraction procedure for the \object{I Zw 1} AGN 
can be deduced from
known flux fractions. By two independent methods,
\citet{1994ApJ...424..627E} and \citet{1998ApJ...500..147S} 
find a stellar fraction of 15\% in the total nuclear K-band  
flux and a stellar fraction of 27\%$\pm$6\% in the total nuclear H-band flux.
Both results refer to a non-extincted QSO component surrounded by a stellar
component which is extincted by 10 to 20 mag.
The first method \citep{1994ApJ...424..627E} is  
based on a comparison of the nuclear NIR colours of
\object{I Zw 1} with the colours of a mean zero-redshift QSO 
\citep{1982MNRAS.199..943H}, while the second method 
\citep{1998ApJ...500..147S} investigates the depths of 
stellar absorption lines in the NIR spectra.
The large stellar flux contribution seems to be connected with the strong 
activity of a $4.5\times 10^7$~yr-old decaying starburst 
\citep{1998ApJ...500..147S}, probably located in the 
circumnuclear molecular ring which has 
recently been resolved in \element[][12]{CO}(1$-$0)
maps \citep{2001IAUS..205..340S} and shows a radius of about 1 kpc.

Since \object{I Zw 1} shows properties of high-redshift QSOs, like the 
blueshifted nuclear emission lines of 
[\ion{C}{iv}]~$\lambda 0.1549\ \mu\mathrm{m} $, 
[\ion{Si}{vi}]~$\lambda 1.961\ \mu\mathrm{m}$, and 
[\ion{Al}{ix}]~$\lambda 2.040\ \mu\mathrm{m}$ 
\citep{1990A&A...240..247B, 1998ApJ...500..147S}, this case study 
implies interesting consequences for 
evolutionary scenarios.
With properties lying inbetween those of QSOs, narrow-line Seyfert~1s, and 
ultraluminous infrared galaxies,
\object{I Zw 1} is discussed as a possible transition object 
from the QSO to the 
ultraluminous infrared stage \citep{2001ApJ...555..719C} regarding the 
original evolutionary scheme for AGN \citep{1988ApJ...325...74S}.
According to its nuclear blue magnitude of $M_\mathrm{B}=-22.5$~mag 
\citep[ adopted for $H_0=75$~km~s$^{-1}$~Mpc$^{-1}$]{1986ApJ...306...64S},
\object{I Zw 1} just belongs to the QSO class 
defined by $M_\mathrm{B} < -22.1$~mag 
\citep[$M_\mathrm{B} < -21.5+5\log h$, ][]{1998yCat.7207....0V}, 
although the host galaxy is visible.
\object{I Zw 1} has a radio-quiet AGN 
\citep{1989ApJS...70..257B, 1995MNRAS.276.1262K} whose radio 
source has not been resolved at the 
observed angular resolutions so far \citep[see][]{1995MNRAS.276.1262K}.
Its  
narrow-line Seyfert 1 characteristics are prototypical of this class, 
like the narrow
H$\beta $ line with a FWHM of 1240~km~s$^{-1}$ 
\citep{1996A&A...305...53B}, the low line ratio 
of [\ion{O}{iii}]/H$\beta$ of 0.49 
\citep{1990AJ.....99...37H}, and the many \ion{Fe}{ii} multiplets 
\citep{1979ApJ...230..360O, 1990AJ.....99...37H, 2001ApJS..134....1V}. 
With an infrared luminosity of $L_\mathrm{IR} = 10^{11.87} L_{\sun} $ 
\citep{2001ApJ...555..719C}, \object{I Zw 1} just fails to be 
an ultraluminous 
infrared galaxy but it still belongs to the luminous infrared objects.

The host of \object{I Zw 1} consists of an extended rotating disk as 
supported by the double-horned line profiles of \ion{H}{i} 
\citep{1985AJ.....90.1642C} and 
CO \citep{1989ApJ...337L..69B}, the spiral arms in the 
optical and NIR
images, and the \element[][12]{CO}(1$-$0) velocity field 
\citep{1998ApJ...500..147S}.
The fuelling mechanisms which are responsible for driving the gas from 
the gas-rich host towards the AGN
are unclear, even on large scales. On the one hand, there is certainly 
no large-scale bar
component as can be seen from the disk images as well as the symmetric
\element[][12]{CO}(1$-$0) isovelocity diagram \citep{1998ApJ...500..147S}.
On the other hand, the blue disk colours of $\mathrm{B}-\mathrm{V}=0.6$--0.7 
\citep{1990AJ.....99...37H} 
indicate enhanced star formation in the disk which can be modelled for 
the north-western spiral arm by either
a young decaying starburst ($1.3\times 10^7$~yr) or an old ($10^{10}$~yr)
constant starburst \citep{1998ApJ...500..147S}. Together with slight tidal
features in the \ion{H}{i} distribution \citep{1999ApJ...510L...7L} and in the 
B'-band image \citep{2001ApJ...555..719C} this suggests that an interaction
with the nearby western companion galaxy may trigger the nuclear activity.
However, the evidence for an interaction is weak \citep{2001ApJ...555..719C}
and it remains
controversial whether \object{I Zw 1} is undergoing a minor merger process.

The new J-band data complement
the knowledge about \object{I Zw 1}, which is already one of
the best-studied QSO hosts.
Observations, data reduction, and data analysis are described in
Sect.~\ref{sec:observations}. The resulting surface-brightness
profile is presented together with the derived gas rotation curve in
Sect.~\ref{sec:profiles}. After an introduction to the applied gravitational
potential models, the surface-brightness profile and the gas rotation
curve are decomposed into structural components 
in Sect.~\ref{sec:decomposition}. The obtained parameters are discussed
in Sect.~\ref{sec:discussion} and compared to the reviewed properties of
the \object{I Zw 1} host. A summary of the results is presented in 
Sect.~\ref{sec:summary}.

\section{Observations and Data Reduction\label{sec:observations}}

NIR imaging data in the J-band have been obtained 
with the ISAAC camera
at the VLT of ESO on Cerro Paranal in Chile. 
All processed data of this paper -- the object frames as well as the standard
star frames for photometric calibration -- were observed within two consecutive
hours. Basic parameters of the observation runs and the corresponding 
air masses are listed in Table~\ref{tbl-1}.

\begin{table*}
\caption{J-band observations with ISAAC at the VLT/UT1.}
\label{tbl-1}
$$
\begin{array}{lccccc}
	\hline
	\noalign{\smallskip}
	\mathrm{Object} & \mathrm{Observation\ Date}   
	& \mathrm{Observation\ Time}   & \mathrm{Detector\ Integration\ Time} &
	\mathrm{Number\ of\ Exposures}  & \mathrm{Air\ Mass} \\
	\noalign{\smallskip}
	\hline
	\noalign{\smallskip}
	\mathrm{I\ Zw\ 1} & 2001-08-19 & 08:05:06$--$08:13:17 
	& 6.00\ \mathrm{s} & 4 & 1.26 \\
	\mathrm{Standard\ star} & 2001-08-19 & 06:14:20$--$06:16:57 
	& 3.55\ \mathrm{s} & 5 & 1.33$--$1.34\\
	\noalign{\smallskip}
	\hline
\end{array}
$$
\begin{list}{}{}
\item[]Characteristics of the ISAAC J-band filter: central wavelength $1.25\ \mu\mathrm{m} $, width $0.29\ \mu\mathrm{m} $. 
\end{list}

\end{table*}

\subsection{Image Reduction and Photometric Calibration\label{sec:photcal}}

Four object frames are 
available in J-band, 
in each of which the object has a different position on the 
detector. 
The images are reduced by means of standard procedures of the 
IRAF software package.
Since \object{I Zw 1} only occupies a small fraction of the total field
of view in an else uncrowded environment, sky and dark current are removed by
subtracting the object frames from each other for all possible 
permutations.
Some modifications of the reduction steps turn out
to optimize the final background noise:
First, the flatfield shows variations of less than 1\% -- measured in several 
$5\times 5$ pixel boxes in the region of interest -- so that a flatfield 
correction is ignored.
Second, the sky-subtracted and aligned images are combined, using a median 
algorithm. The advantage of this algorithm over the average algorithm is that
residual stars in the sky-subtracted images are effectively
removed during combination.

A standard star from the catalogue of the faint NIR LCO/Palomar
NICMOS standards \citep{1998AJ....116.2475P} was observed for photometric 
calibration. 
The match between these standards 
and the ISAAC filters has not yet been experimentally
verified but the colour terms are expected to be close to zero.
Observed in the same acquisition modes, the calibration frames
are reduced parallel to the object frames.
The flux
of the star is measured at all five different detector positions in the 
unaligned frames to account for detector variations. 
Mean and standard deviation of the five values are used for the final 
zero-point determination and the calibration
error, respectively. 
The derived zero-point is corrected for air mass with the canonical 
average value of
$0.11\ \mathrm{mag}\ (\mathrm{air\ mass})^{-1}$ 
given for J in the ISAAC manual.

\subsection{Subtraction of the Nucleus}

In order to investigate the host galaxy of \object{I Zw 1}, 
the non-stellar brightness contribution of the QSO has to be subtracted. 
For a two-dimensional subtraction, the PSF of the original J-band image 
is built with
a gaussian 
using the IRAF task PSF. To account for asymmetries, the residuals of the fit
are stored in an additional lookup table. In order to subtract only the 
non-stellar portion of nuclear flux, the PSF has to be appropriately scaled.
Without any prior assumptions about bulge shape or fitting constraints, the 
scaling is determined from the known stellar contributions to the
total nuclear H- and K-band fluxes 
\citep{1994ApJ...424..627E, 1998ApJ...500..147S} as mentioned in 
Sect.~\ref{sec:intro}. 
The corresponding J-magnitude is computed using the JHK colours of a 
mean zero-redshift QSO \citep{1982MNRAS.199..943H}.
The QSO turns out to be 0.08~mag fainter than the total nuclear J-magnitude 
starting from the K-band estimate and 0.19~mag fainter
starting from the H-band estimate. As expected, 
both results are similar and
the mean value of 0.13~mag is used for the final scaling of the PSF.

\subsection{Deriving the Surface-Brightness Profile}

The analysis of the \object{I Zw 1} host is restricted 
to a one-dimensional fitting of 
radial profiles. A one-dimensional decomposition turns out to be 
sufficient for the goals of this NIR study, since (i) spiral arms are not as 
pronounced in the NIR as in the optical and (ii) \object{I Zw 1} is seen 
nearly face-on so that its brightness distribution is radially symmetric 
to a good approximation -- the latter being supported by a previous 
NIR investigation of \object{I Zw 1} 
\citep{1999ApJS..125..363P} in which ellipse fitting resulted in an  
almost constant ellipticity of about 0.1 throughout the disk. 
The radial surface-brightness profile is determined by measuring the flux per 
arcsec$^2$ along circles around the centre of \object{I Zw 1} 
using the IRAF task PRADPROF.
Error bars are computed as the standard deviation of 
numerous values along neighbouring circles in radially logarithmic bins.
Neglecting the slight ellipticity of the isophotes adds radial 
smearing to the measured profile. To estimate the amount of smearing, 
the radially 
averaged profile is cross-checked with a direct cut along the kinematic
major axis. Both curves agree well within the error bars.

Since the minimum integration time of the ISAAC detector was increased by a 
factor of 2 on July 24th, 2001, the nuclear J-band flux in a few central pixels
exceeds the analog-to-digital-units (ADU) level above 
which non-linearity of the array occurs. 
To make sure that no nuclear flux
is lost in the J-band images, the formula
\begin{equation}
f_{\mathrm{true}}=
f_{\mathrm{measured}}+4.75\times 10^{-11} f_{\mathrm{measured}}^3,
\end{equation}
provided in the ISAAC reduction guide,
is used to correct the flux in the
brightest pixels. 
The difference between the corrected and the uncorrected values
is expressed as errors of the data points.

According to the relative photometric calibration based on the 
LCO/Palomar NICMOS system
\citep{1998AJ....116.2475P}, the surface brightness is given in units of
mag~arcsec$^{-2}$. 
In order to obtain an absolute calibration in units of 
Watt~m$^{-2}$~Hz$^{-1}$, the UKIRT absolute flux calibration 
\citep{1976ApJ...208..390B} is used, taking into account the 
slight zero-point offset of 
$J_\mathrm{NICMOS}-J_\mathrm{UKIRT}=0.034\pm 0.004 $ \citep{2001MNRAS.325..563H}.  
The resulting conversion formula is given by
\begin{equation}
S = 
10^{-26}\times 10^{\epsilon _\mathrm{J} - 0.4 m},
\end{equation}
where $S$ is the flux, $m$ the magnitude, and $\epsilon _\mathrm{J}=3.217 $
the absolute calibration factor already including the zero-point offset.
The luminosity $L$ follows from $L=4\pi D^2 \times S$, 
where $D$ is the distance of \object{I Zw 1}.
For the final calibration in solar luminosities, the J-band luminosity of 
the sun is taken from
Table~2.1 in \citet{1998gaas.book.....B} and is divided 
by the band width of the filter.
In total, the applied transformation formula is
\begin{equation}
L=
2.0066\times 10^{16} \times 10^{-0.4m} \times L_{\sun}.
\end{equation}

\section{Results\label{sec:profiles}}

\subsection{The J-Band Image}

\begin{figure}
	\resizebox{\hsize}{!}{\includegraphics{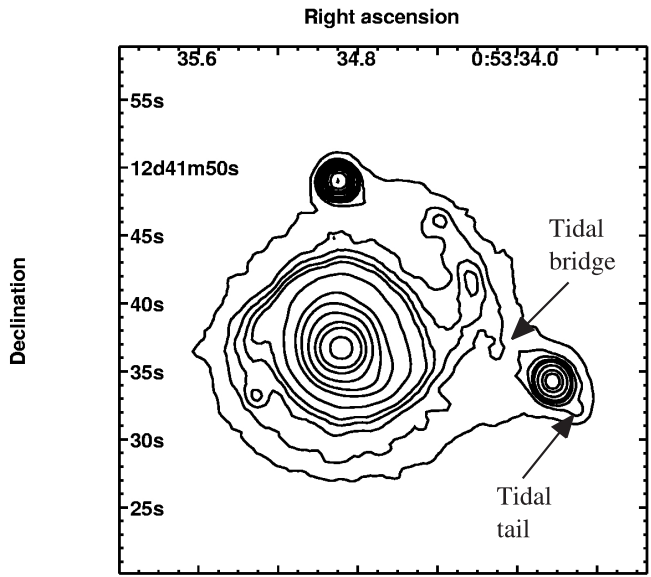}}
	\caption{Contour J-band image of the \object{I Zw 1} host. 
	The displayed contour levels are 21.16, 20.63, 20.38, 20.16, 19.97, 
	19.51, 18.83, 18.41, 17.88, 17.25, 16.64, and 14.65 
	mag arcsec$^{-2}$. The position of the tidal bridge and the tidal
	tail in the interaction zone is indicated.}
	\label{fig:image}
\end{figure}

The typical characteristics of the \object{I Zw 1} environment -- like the
two-armed spiral structure, the northern foreground star, and
the nearby western galaxy -- can be seen in the 
reduced and calibrated J-band image 
(Fig.~\ref{fig:image}). 
The image gives strong evidence
for an interaction between \object{I Zw 1} and the companion, since it allows 
an unprecedented view on the possible interaction zone between
the western spiral arm and the companion, showing an elongation of the
companion as well as two features which suggest 
a tidal bridge and a tidal tail. The detailed 
insights are provided by the high
angular resolution -- with a 
Gaussian full width at half maximum (FWHM) of 
$0\farcs 6$ of the stellar PSF -- and the high sensitivity -- with a 
background noise of 
0.24 ADU~s$^{-1}$~pixel$^{-1}$, corresponding to a limiting surface 
brightness of 22.5~mag~arcsec$^{-2}$ at a 1$\sigma $ level, 
or a limiting surface brightness 
of 21.3~mag~arcsec$^{-2}$ at a 3$\sigma $ level -- of the ISAAC observations.

\subsection{The Surface-Brightness Profile\label{sec:freeman}}

\begin{figure}
	\resizebox{\hsize}{!}{\includegraphics{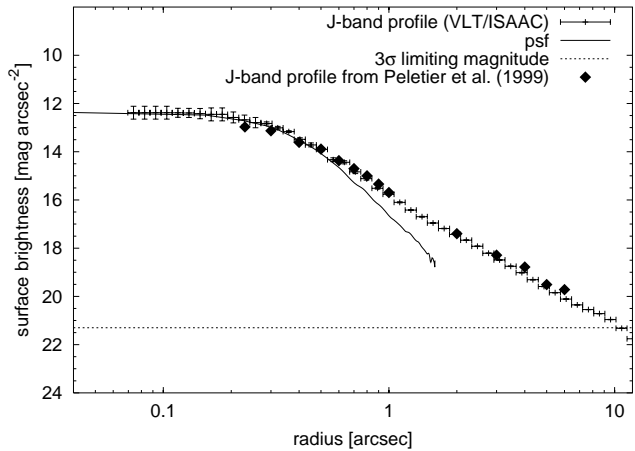}}
	\caption{Comparison between the VLT/ISAAC J-band 
	surface-brightness profile
	of \object{I Zw 1}, corrected for detector non-linearity, 
	and earlier data from 
	\citet{1999ApJS..125..363P}. Both profiles show a 
	good agreement, even within 
	the FWHM of the PSF ($0\farcs 6$) of the ISAAC observations. 
	The 3$\sigma $ limiting surface brightness of the ISAAC profile is 
	also shown.}
	\label{fig:comparepeletier}
\end{figure}

The derived J-band surface brightness profile is presented in 
Fig.~\ref{fig:comparepeletier} together with the PSF, 
as measured for a nearby star,
and the $3\sigma $ limiting surface brightness. 
The profile is compared to the J-band data 
previously measured by \citet{1999ApJS..125..363P}. Both curves agree well, 
even within the 
FWHM of the PSF of the ISAAC observations.

\begin{figure}
	\resizebox{\hsize}{!}{\includegraphics{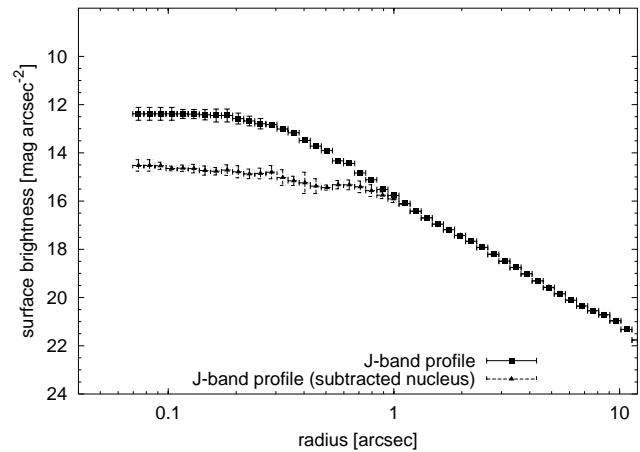}}
	\caption{Comparison of the original J-band surface-brightness profile
	of the \object{I Zw 1} host and the profile from which the QSO 
	nucleus is subtracted. 
	}
	\label{fig:comparesubnuc}
\end{figure}

\begin{figure}
	\resizebox{\hsize}{!}{\includegraphics{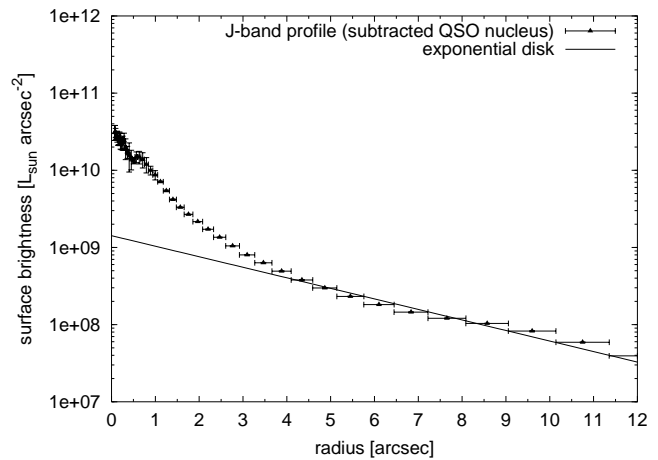}}
	\caption{J-band surface-brightness profile of 
	\object{I Zw 1} calibrated in solar
	luminosities. An exponential disk component is fitted to emphasize the 
	Freeman type I of the profile.}
	\label{fig:freeman}
\end{figure}

The original profile and the QSO-subtracted profile 
of the \object{I Zw 1} host are compared in 
Fig.~\ref{fig:comparesubnuc}.
Although the subtraction method does not rely on a certain shape of the
host bulge, the residual stellar component shows a flat light 
distribution within the error bars. 
An exponential disk is fitted to 
the calibrated and QSO-subtracted surface-brightness profile in 
Fig.~\ref{fig:freeman} in order to model the clearly evident extended 
rotating disk of \object{I Zw 1}.

This fit shows that an additional bulge component is needed to describe the 
whole profile consistently.
Bulge and disk components can be clearly distinguished, 
as it is typical of Freeman type I galaxies \citep{1970ApJ...160..811F}.
The host belongs to the high-surface-brightness galaxies which, per
definition, have a central blue disk surface brightness of
$\mu _0 (\mathrm{B})< 21.65$~mag~arcsec$^{-2}$ \citep{1970ApJ...160..811F}.
The exponential disk of \object{I Zw 1} 
has a central surface brightness of about 
$\mu _0(\mathrm{J})=1.4\times 10^9 L_{\sun} $~arcsec$^{-2}$ or 
$\mu _0(\mathrm{J})=
17.9$~mag~arcsec$^{-2}$. Assuming that the central part of the
galaxy shows almost the same colours as the outer parts of the disk, this 
corresponds to 
$\mu _0(\mathrm{B})\approx (19.5\pm 0.5)$~mag~arcsec$^{-2}$, which is 
brighter than the canonical Freeman value. 
Here, a $\mathrm{B}-\mathrm{J}$ of 
$1.6\pm 0.5$ for the \object{I Zw 1} host is used as 
derived from the broad band spectrum
in Fig.~9 of \citet{1998ApJ...500..147S} and the $\mathrm{B}-\mathrm{V}$ 
colour of 0.6 to 0.7 
\citep{1990AJ.....99...37H}. 
The high central disk surface brightness of the \object{I Zw 1} host 
reflects the finding
by \citet{1999ApJ...510..637H} that exponential disks of 
Seyfert galaxies tend to 
be brighter than normal spirals. With 
$\mu _0(\mathrm{B})\approx (19.5\pm 0.5)$~mag~arcsec$^{-2}$ 
the \object{I Zw 1} disk is 
significantly brighter than the mean B-band surface brightness of
21.44~mag~arcsec$^{-2}$ of normal spirals shown in Fig.~6 of 
\citet{1999ApJ...510..637H}.

\subsection{The Gas Rotation Curve \label{sec:rotcurve}}

\begin{figure}
	\resizebox{\hsize}{!}{\includegraphics{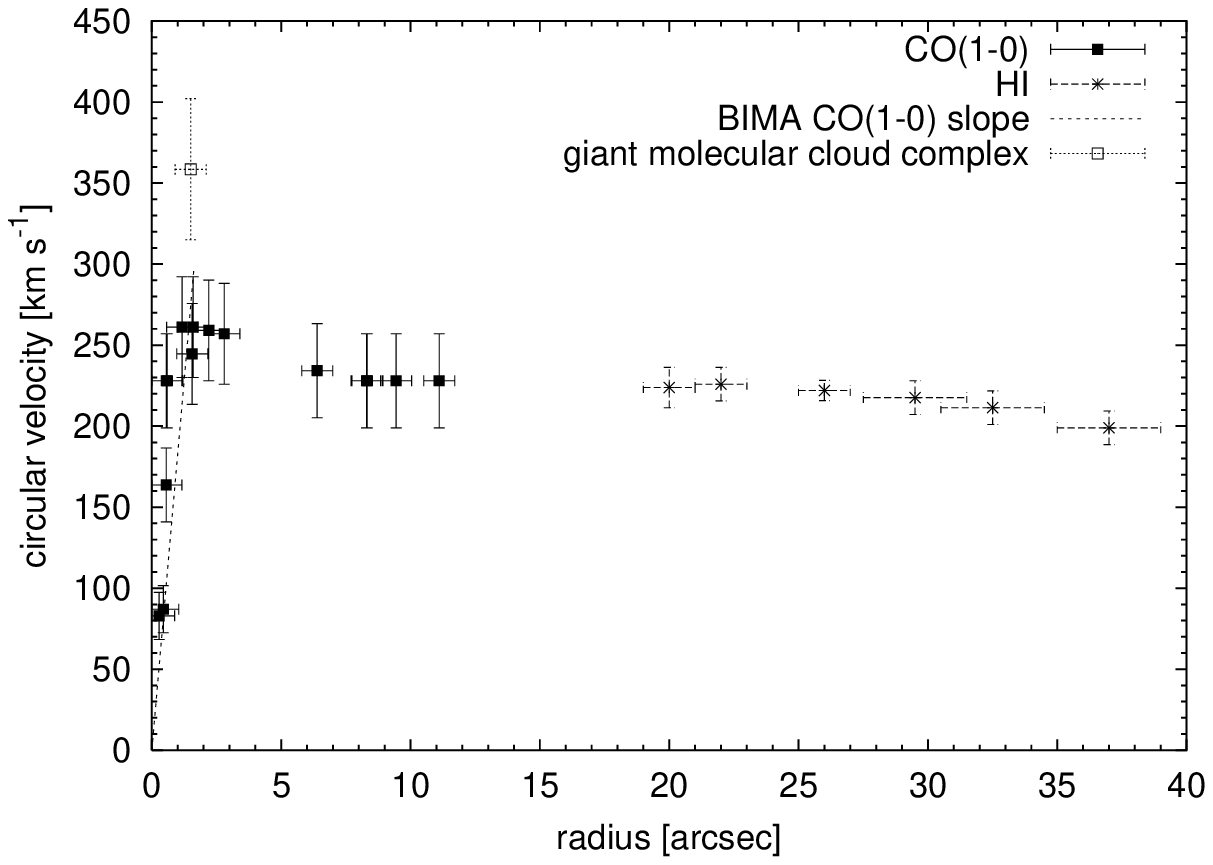}}
	\caption{Rotation curve of \object{I Zw 1} corrected for a constant 
	inclination of $38\degr $. The inner slope of the 
	\element[][12]{CO}(1$-$0) data from observations with
	the Plateau de Bure interferometer by \citet{1998ApJ...500..147S}, 
	(filled squares), is compared with $0\farcs 7$ resolution BIMA 
	observations by \citet{2001IAUS..205..340S}, (dashed line). 
	Towards larger radii, the curve is
	extended on the basis of \ion{H}{i} data from VLA measurements by 
	\citet{1999ApJ...510L...7L}, 
	(asterisks). The open square marks the velocity of a 
	single giant molecular
	cloud complex on the south-eastern side of the \object{I Zw 1} 
	nucleus.}
	\label{fig:rotcurve}
\end{figure}

The composite rotation curve, plotted in 
Fig.~\ref{fig:rotcurve} together with 5$\sigma $ error bars, 
is corrected for the \object{I Zw 1} disk inclination of $(38\pm 5)\degr $ 
under the assumption of purely circular gas motions.
Gas velocities at outer radii are taken from a VLA \ion{H}{i} map 
\citep{1999ApJ...510L...7L}. 
These observations have an angular resolution of about $18\farcs 6$.
For the inner rotation curve, 
gas velocities are derived from \element[][12]{CO}(1$-$0) maps 
\citep[ Fig.~5]{1998ApJ...500..147S}, measured with the Plateau de Bure
interferometer at angular resolutions of $1\farcs 9$ and $5\arcsec $.
The inner slope
is verified by a comparison with \element[][12]{CO}(1$-$0) BIMA 
observations taken at a 
higher resolution of about $0\farcs 7$ \citep{2001IAUS..205..340S}. 
The good agreement of the different observations indicates that
the inner rise of the rotation curve is not yet affected by the mass 
concentration of the active nucleus at these resolutions. The single 
deviating data point corresponds to an isolated peak in the \element[][12]{CO}(1$-$0)
position-velocity diagrams \citep{1998ApJ...500..147S} and is interpreted as
a giant molecular cloud complex by the same authors. The deviation from the
general circular velocity at this radius corresponds to a cloud orbit which is
inclined by $20\degr $ with respect to the \object{I Zw 1} disk. 

Like the surface-brightness profile, the gas rotation curve supports the
classification of the \object{I Zw 1} host as a high-surface-brightness 
galaxy. It shows
the characteristic steep increase at the inside and slight decrease at the
outside found for the
universal rotation curve of high surface-brightness galaxies by 
\citet{1996MNRAS.281...27P}. 
The overall shape of the rotation curve of the \object{I Zw 1} host is also 
similar to the 
rotation curve of the Milky Way 
\citep[ Fig. 1]{1999ApJ...523..136S} and
agrees with the mean rotation
curve, presented for galaxies with disk rotation velocities between 
200~km~s$^{-1}$ and 250~km~s$^{-1}$ in the same article (Fig.~5). According
to the authors, the central peak, the broad disk component, and the 
high velocities
at large radii characterize the \object{I Zw 1} host as a massive 
Sb or Sc galaxy. 
A similar classification of the \object{I Zw 1} host as a Sc galaxy is given by
\citet{2000ApJS..126..271J}.

\subsection{Decomposition\label{sec:decomposition}}

In order to derive structural information about the underlying galaxy
of the \object{I Zw 1} QSO, the host is decomposed
into contributions from bulge, disk, and dark halo components. 
Taking one potential model for each component, the one-dimensional 
decomposition has 
six free parameters, i.e. the scale lengths and the M/L ratios 
of the bulge and disk
potentials as well as, where required, the scale length and mass 
contribution of a dark halo.  
The bulge and disk scale lengths which are used to model the
rotation curve are derived from the distribution of luminous matter as 
traced by the surface brightness profile.
This approach results in mean M/Ls for the bulge and disk components and
is straightforward for the purpose of discussing 
general trends in M/L.
However, the reader should be aware that M/Ls -- even in the NIR --
are local observables which vary due to effects of dust and stellar 
populations. 
Decompositions of the rotation curve are usually not unique 
\citep[e.g.][]{1986RSPTA.320..447V, 2002A&A...388..793B} so that 
four different solutions will be presented and discussed.
All fits are performed with an 
implementation of the non-linear 
least-squares method.

\subsubsection{Potential Models}

The rotation curve and the surface-brightness profile are decomposed with a
Plummer model for the bulge component and a thin-disk Kuzmin model for the 
disk component. These models are chosen due to mathematical 
convenience, since both are consistently described by the Miyamoto-Nagai 
gravitational potential \citep{1975PASJ...27..533M}
\begin{equation}
\Phi_i (R,z)=
\frac{-GM_i}
	{\sqrt{R^2+\left(a_i+\sqrt{z^2+b_i^2}\right)^2}}, 
\end{equation}
where $M_i$ is the mass of component $i$, $a_i$ and $b_i$ are two scale 
lengths, and G is the gravitational constant.
The case $a_i=0$ applies to the Plummer model, the case $b_i=0$ to the Kuzmin 
disk.
The total potential of both components is the superposition
\begin{equation}
\Phi (R,z)=
\frac{-GM_{\mathrm{P}}}
	{\sqrt{R^2+z^2+b_{\mathrm{P}}^2}}+
\frac{-GM_{\mathrm{K}}}
	{\sqrt{R^2+\left(z+a_{\mathrm{K}}\right)^2}}, 
\end{equation}
where P denotes the Plummer parameters and K the Kuzmin parameters.
For the rotation curve fits, another Plummer potential is added to account
for a dark halo component.
The circular velocities $v_c(R)$ follow from 
\begin{equation}
v_c^2(R)=
\sum _i \left( R\frac{\partial \Phi _i}{\partial R} \right)_{z=0} .
\end{equation}
For Plummer and Kuzmin potentials this means explicitly
\begin{equation}
v_c^2(R)=
\frac{R^2GM_{\mathrm{P}}}
	{\left(R^2+b_{\mathrm{P}}^2\right)^{\frac{3}{2}}}+
\frac{R^2GM_{\mathrm{K}}}
	{\left(R^2+a_{\mathrm{K}}^2\right)^{\frac{3}{2}}}.
\end{equation}
The corresponding total surface density $\Sigma (R) $ is given by
\begin{equation}
\Sigma (R) = 
\frac{b_{\mathrm{P}}^2M_{\mathrm{P}}}
	{\pi \left(R^2+b_{\mathrm{P}}^2\right)^2}+ 
\frac{a_{\mathrm{K}} M_{\mathrm{K}}}
	{2\pi \left(R^2+a_{\mathrm{K}}^2\right)^{\frac{3}{2}}}.
\label{eq:massdens}
\end{equation}
Assuming a constant mass-to-light ratio for 
each component, the same formula applies to the surface brightness, 
if masses are substituted by luminosities.

\subsubsection{Decomposition of the Surface-Brightness Profile}

As the \object{I Zw 1} host is a Freeman Type I galaxy 
(Sect.~\ref{sec:freeman}), its profile can best be described by a combination
of bulge and disk component.

\begin{figure}[h]
	\resizebox{\hsize}{!}{\includegraphics{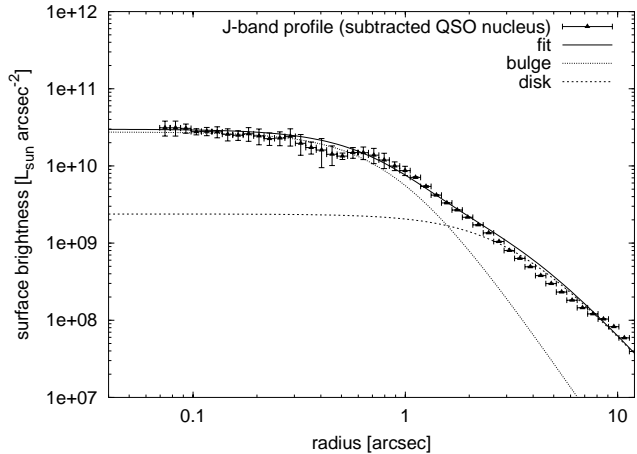}}
	\caption{Final model for the surface-brightness 
	profile. The contributions of the Plummer bulge and 
	the Kuzmin disk are also visualized. In order to resolve the 
	bulge region better, radial distances are plotted logarithmically.}
	\label{fig:surfbright_fit}
\end{figure}

A decomposition of the J-band surface-brightness profile, consisting of the 
two-component model with Plummer bulge and Kuzmin disk, is shown 
in Fig.~\ref{fig:surfbright_fit}. In this decomposition, 
the Kuzmin model is fitted to that part of the profile where 
possible residua of an unclean
QSO subtraction do certainly not bias the profile shape. The remaining
surface brightness is ascribed to the Plummer bulge.
The decomposition defines the scale lengths of bulge and disk as
$0\farcs 90$ and $3\farcs 08$, respectively, which corresponds
to a bulge-to-disk scale length ratio of about 0.29.
The luminosity of the bulge results in  
$7.05\times 10^{10}\ L_{\sun}$ and is roughly half as large 
as the luminosity
of the disk of $14.21\times 10^{10}\ L_{\sun}$. 

\subsubsection{Decomposition of the Gas Rotation Curve and Dynamical Mass-to-Light Ratios}

\begin{figure*}
\subfigure[]{
	\resizebox{0.5\hsize}{!}
	{\includegraphics{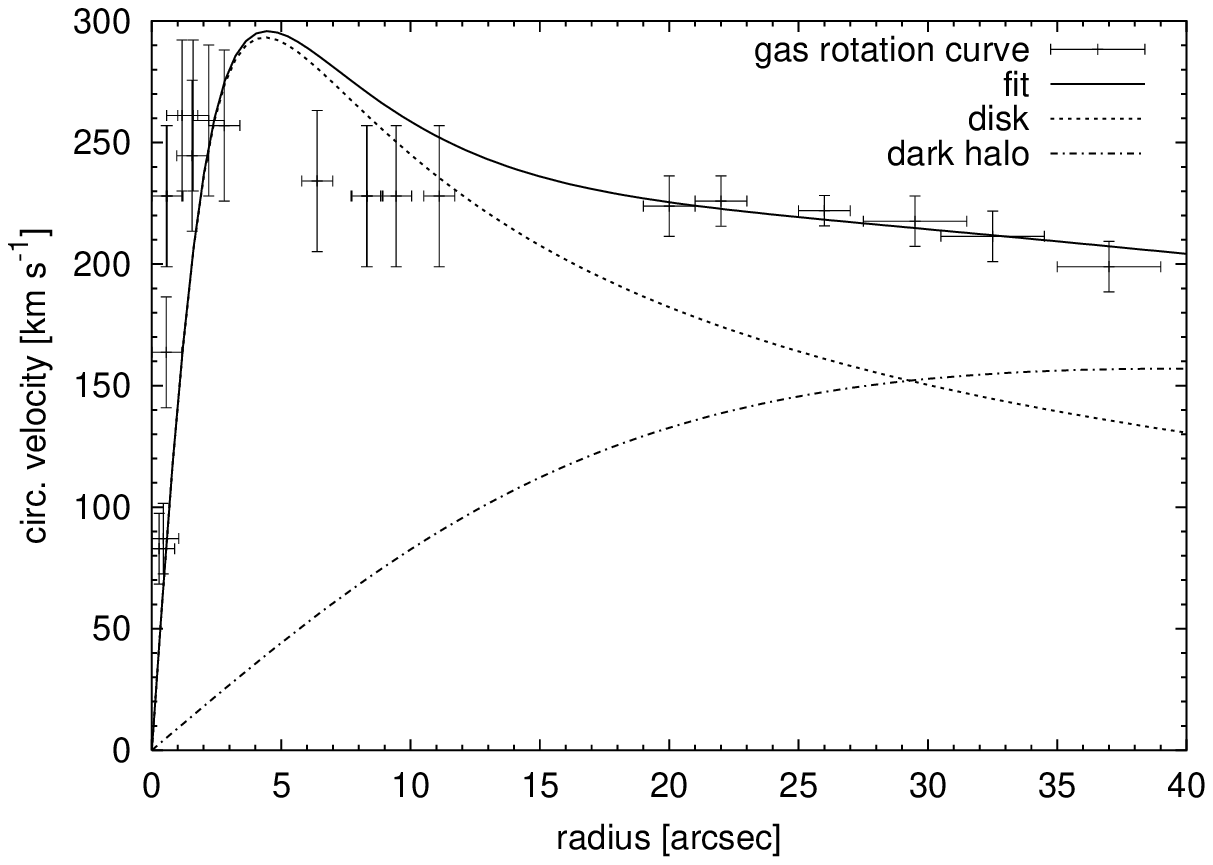}}}
\subfigure[]{
	\resizebox{0.5\hsize}{!}
	{\includegraphics{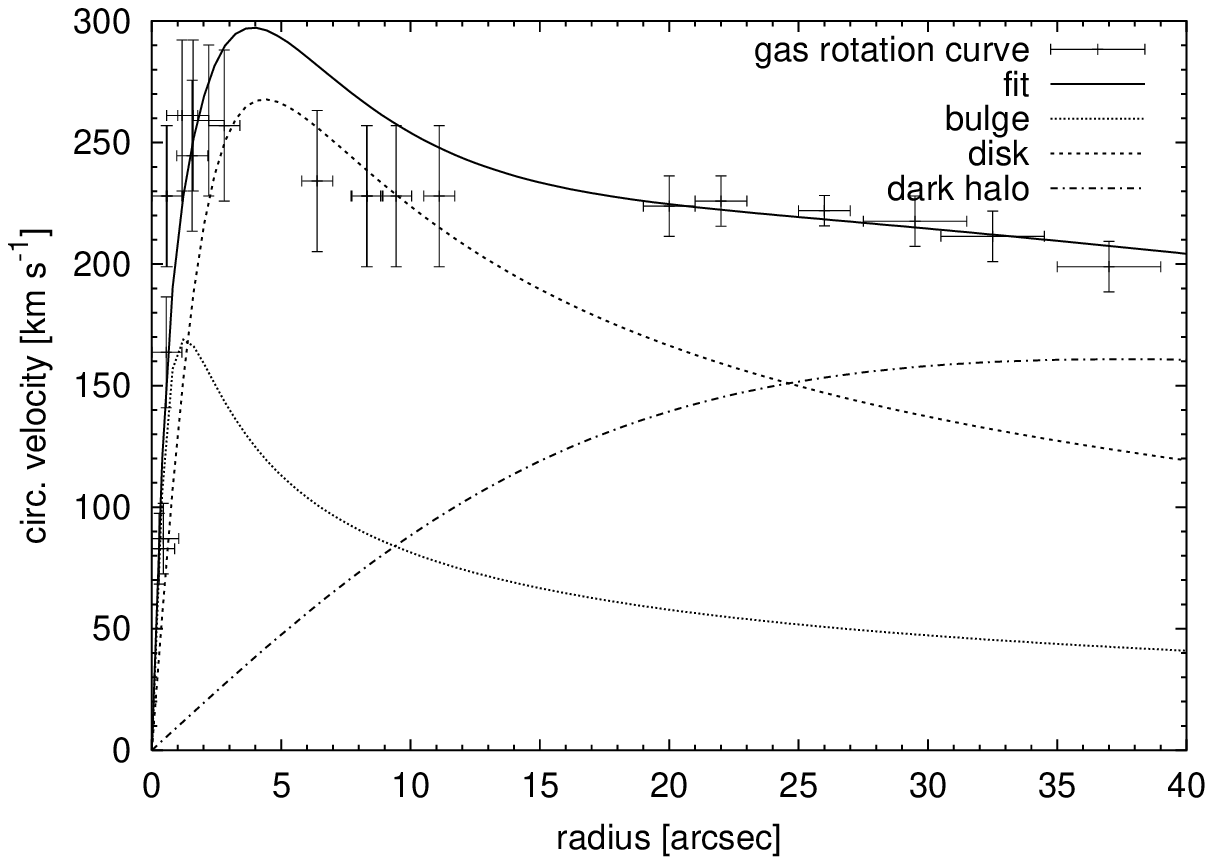}}}
\subfigure[]{
	\resizebox{0.5\hsize}{!}
	{\includegraphics{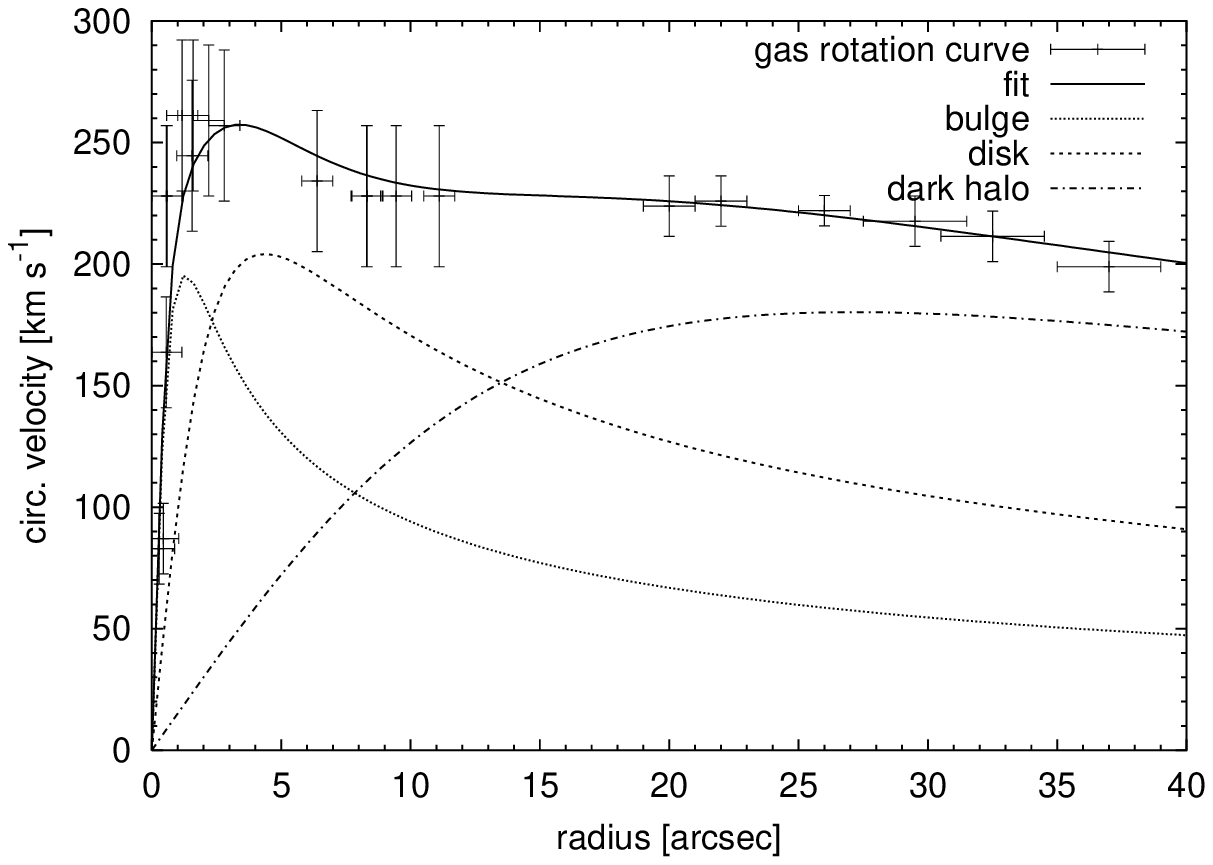}}}
\subfigure[]{
	\resizebox{0.5\hsize}{!}
	{\includegraphics{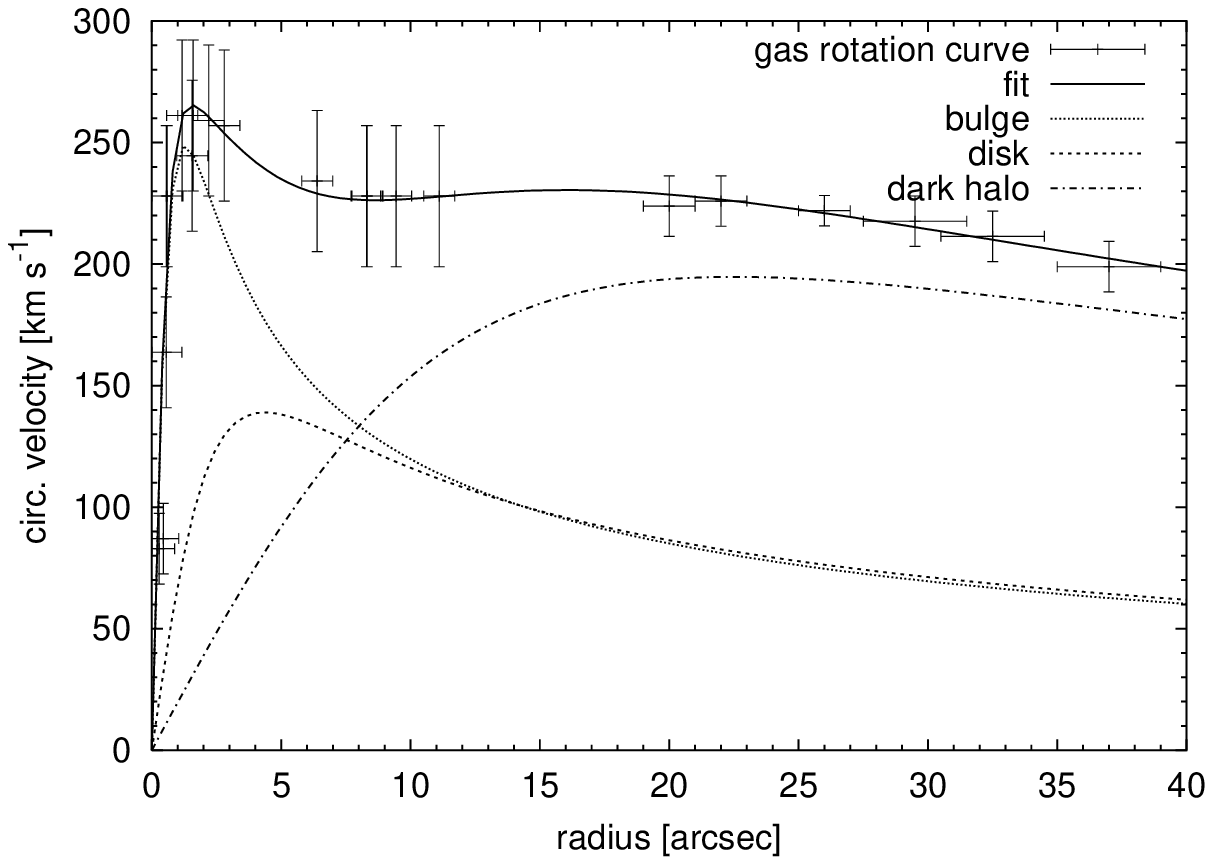}}}
	\caption{Four examples of a decomposition of the \object{I Zw 1} 
	gas rotation curve.
	The model curve (solid line) is shown together with the individual 
	contributions of bulge (dotted line), disk (dashed line), and dark halo
	(dash-dot line). \textbf{a)} Fit with only 
	a disk and a dark halo (\emph{disk only}). \textbf{b)} 
 	Maximum-disk fit (\emph{max disk}). 
	\textbf{c), d)} All fitting parameters are unconstrained and 
	fitted simultaneously.
	The left panel shows a case in which more weight is put 
	on the steep increase at the inside
	of the rotation curve by confining the lower limit of
	the fitting region to $0\farcs 4$ (\emph{free a}). 
	In the right panel 
	the fitting region starts at $0\farcs 6$ (\emph{free b}). 
	See text for a detailed description.}
	\label{fig:rotfit1}
\end{figure*}

Fig.~\ref{fig:rotfit1} 
shows four decomposition models for the
gas rotation curve of \object{I Zw 1}. 
Each fit is obtained under different contraints
for the model components.
The final parameters are listed in Table~\ref{tab:rotcurve_fit}. The 
resulting $\chi ^2$ is added for these cases in which the fitting 
procedure allows a specification.

\begin{table*}
\caption{Parameters of the decomposition. The table lists the 
$\chi ^2$ values of the four different fitting methods (Col. 2), 
the bulge M/L (Col. 3), the disk M/L (Col. 4), 
the halo mass (Col. 5), the halo scale 
length (Col. 6), the maximum circular velocity of the disk component 
expressed
as a percentage of the maximum disk circular velocity (Col. 7), 
and the X parameter of the 
respective disk component (Col. 8). The scale lengths of bulge and disk are given by the
decomposition of the surface brightness profile as $0\farcs 90$ and 
$3\farcs 08$, respectively. The meaning of the X parameter is explained
in the text (Sect. \ref{sec:swingampli}).}
\label{tab:rotcurve_fit}
$$
\begin{array}{lccccccc}
	\hline
	\noalign{\smallskip}
	\mathrm{Fit} & \chi ^2 
	& \mathrm{(M/L)_{bulge}}\ [M_{\sun}/L_{\sun} ]   
	& \mathrm{(M/L)_{disk}}\ [M_{\sun}/L_{\sun}] 
	& \mathrm{M_{halo}}\ [10^{10}\ M_{\sun}] 
	& \mathrm{Scale\ Length_{halo}\ [\arcsec ]}   
	& \mathrm{\%} 
	& X \\
	\noalign{\smallskip}
	\hline
	\noalign{\smallskip}
	$\emph{disk only}$  &  3.50  &  -     &  1.33  &  49.18  &  28.00  &
	-   & -   \\
	$\emph{max disk}$   &  -     &  0.26  &  1.11  &  49.03  &  26.58  &
	100 & 0.9 \\
	$\emph{free a}$     &  1.51  &  0.34  &  0.64  &  44.38  &  19.17  &
	76  & 1.5 \\
	$\emph{free b}$     &  0.11  &  0.56  &  0.30  &  43.02  &  15.91  &
	52  & 3.2 \\
	\noalign{\smallskip}
	\hline
\end{array}
$$
\end{table*}

As a cross-validation of the two-component decomposition of the 
surface-brightness profile,
Fig.~\ref{fig:rotfit1}~a) shows a fit in which
only a disk and a dark halo model are used while  
the bulge component is left out (\emph{disk only}). 
The high $\chi ^2$ indicates
that the peak of the rotation curve at small radii 
cannot sufficiently be described by
such a model. In order to obtain a good fit of the rotation curve with only
a disk and dark-matter model, the disk scale length would have to be chosen 
very small ($\sim 1\farcs 37$). Such a Kuzmin disk, however, does not yield
a good fit of the surface-brightness profile at large radii.

Fig.~\ref{fig:rotfit1}~b) presents a maximum-disk fit 
(\emph{max disk}) with bulge, disk, and dark halo component.
The four 
parameters (Table~\ref{tab:rotcurve_fit}) are determined by the 
following procedure: First, only the disk M/L is adjusted to the rotation
curve in order to obtain a maximum contribution of the disk component.
Bulge and dark halo are added in a second and third step to finetune the
total velocity profile.
Since the 
fitting is done in individual steps, it is not possible to 
give a general $\chi ^2$ for this procedure.

The lower panels of Fig.~\ref{fig:rotfit1} show two least-squares 
minima for a non-maximum-disk fit in which all four parameters
(Table~\ref{tab:rotcurve_fit}) are fitted simultaneously.
In Fig.~\ref{fig:rotfit1}~c), 
more weight is put on the steep increase
of the rotation curve at small radii 
by giving a lower limit of $0\farcs 4$ for the fitting
region (\emph{free a}) whereas the fit in Fig.~\ref{fig:rotfit1}~d) 
is confined to a region starting at $0\farcs 6$ (\emph{free b}).
The first procedure results in a disk component 
whose circular velocities amount to 76\% of the corresponding circular 
velocities of the maximum disk. 
The second procedure yields a more massive bulge
component and a dominant dark matter contribution within the disk region of the
\object{I Zw 1} host. Here, only 52\% of the maximum disk circular velocities
are attributed to the disk component.

\section{Discussion\label{sec:discussion}}

\subsection{Distribution of Luminous Matter in the \object{I Zw 1} Host}

By comparing the decompositions of the 
J-band surface-brightness profile and the gas rotation
curve it becomes obvious that at least two components of 
luminous matter -- a bulge and a disk -- are necessary to 
model the \object{I Zw 1} host. The surface-brightness fit results in a 
bulge and a
disk with almost equal luminosities and a bulge-to-disk scale length ratio 
of 0.29.
This is in good agreement with the relations presented by 
\citet{2000A&A...360..431M} for nearby AGN spiral hosts, who found a 
bulge-to-disk scale length ratio of about 0.2.

\subsection{Mass Distribution and M/Ls\label{sec:swingampli}}

\begin{figure}
	\resizebox{\hsize}{!}{\includegraphics{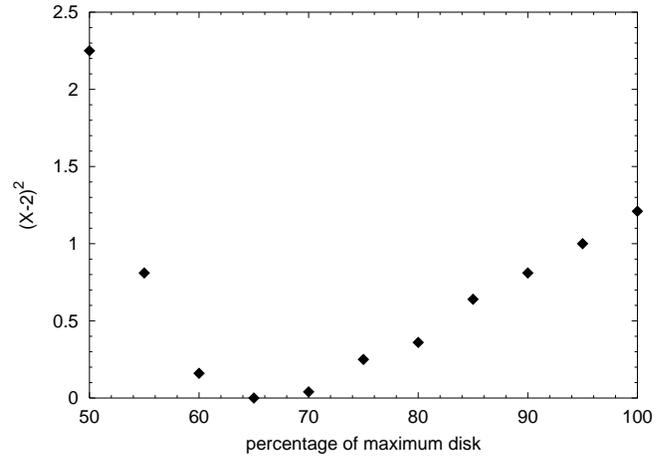}}
	\caption{Deviation of the X parameters of various 
	disk solutions from the X parameter 
	($ X\approx 2 $) at which amplification is most effective.
	The deviation is expressed as the quadratic deviation 
	$ (X-2)^2 $ 
	from $ X=2 $. The disk solutions are expressed in terms of their
	maximum circular velocities as a percentage of the maximum disk.}
	\label{fig:Xparam}
\end{figure}

The maximum-disk assumption 
\citep{1986RSPTA.320..447V} is often applied to have a
criterion for unique rotation curve fits.
Maximum-disk fits seem to provide satisfying
models for Freeman type I as well as Freeman type II galaxies
\citep{2000AJ....120.2884P}, although this may only be true for 
high-surface-brightness galaxies \citep{1997MNRAS.290..533D}. As a 
high-surface-brightness galaxy of Freeman type I, the maximum-disk fit should,
therefore, provide a good estimation of the M/Ls of the \object{I Zw 1} 
host components. 
However, \citet{1993A&A...275...16B} argues that, for realistic disk 
scale heights, stellar velocity dispersion only allows disk components 
with circular velocities of about 63\% of the observed maximum velocities.
Such a Bottema disk is close to the two results
presented in Figs.~\ref{fig:rotfit1}~c) and \ref{fig:rotfit1}~d) 
with disk components of 76\% and 52\% of the maximum disk. 
\object{I Zw 1} is an interacting galaxy so that the observed two-armed
spiral structure could be driven by the companion. However, assuming 
swing amplification 
\citep[see][]{1981seng.proc..111T, 1984PhR...114..321A, 2001A&A...368..107F} 
as the underlying mechanism for the 
amplification of the \object{I Zw 1} spiral, the disk surface density $\Sigma $
at radius $R$ can be obtained from the relation
\begin{equation}
\Sigma =
\frac{R \kappa ^2}{2 \pi G X m},
\label{eq:swingampli} 
\end{equation}
where $\kappa $ is the epicyclic frequency, $G$ the gravitational constant, 
$m$ the number of spiral arms, and $X$ a parameter measuring the local ratio
of disk mass to total mass. For the flat or slightly falling rotation curve
of the \object{I Zw 1} host, amplification is most effective at an 
X parameter of about 2 \citep{1981seng.proc..111T}.
For the given disk scale length of the \object{I Zw 1}
host of $a_{\mathrm{K}}=3\farcs 08$, the circular velocity of the 
Kuzmin disk peaks at
$\sqrt{2} a_{\mathrm{K}} \approx 4\farcs 36$ 
where 
$ \kappa \approx 73.9 $~km~s$^{-1}$~arcsec$^{-1} $ can be   
taken from the rotation curve. 
Using equation~(\ref{eq:swingampli}), the deviation of the X parameters 
from the X parameter of $ \sim 2 $ at which amplification is most 
efficient is plotted for various disk solutions in 
Fig.~\ref{fig:Xparam}. As above, the disks are defined by their respective 
maximum circular velocities expressed as a percentage of the maximum
disk velocity.
Fig.~\ref{fig:Xparam} shows that in the case of \object{I Zw 1} amplification
is most effecient for a $ \sim 65\% $ disk.
For the disks in fits \emph{max disk} (100\%) and \emph{free b} (52\%) 
amplification is already rather inefficient (Table~\ref{tab:rotcurve_fit})
with X parameters of 0.9 
and 3.2, respectively.
Thus, the fits \emph{max disk} and \emph{free b} are limiting
cases so that the dynamical J-band M/Ls and their errors are estimated as 
the mean and the deviations of both solutions, resulting in  
$(0.4\pm 0.2)\ M_{\sun} / L_{\sun}$ for the bulge and 
$(0.7\pm 0.4)\ M_{\sun} / L_{\sun}$ for the disk.

\subsection{Effects of Gas Masses and Dust Extinction}

The discussion so far has only concentrated on dynamical M/Ls of the 
\object{I Zw 1} host components, ignoring any effects of gas contributions and
dust extinction. The real stellar M/Ls, however, are certainly lower
than the given values, since the mass 
contributed by gas has to be subtracted from the measured 
dynamical mass and the 
light absorbed by dust has to be added to the measured luminosity. 
With at 
maximum 7\% of total cold gas mass \citep{1994ApJ...424..627E}, 
the mass correction for the disk component is negligible and within
the error bars. 
As the disk is seen nearly face-on, the extinction correction
is also assumed to be low with a maximum mean J-band extinction of 0.5 mag.
The effects on the M/L of the bulge are probably much stronger.  
With a gas fraction of 20\% in the bulge 
\citep{1998ApJ...500..147S}, the 
stellar bulge M/L amounts to 80\% of the dynamical bulge M/L.
The visual extinction of the stellar 
component in the circumnuclear region is reported to be about 10 mag
\citep[][ and Sect. \ref{sec:intro}]{1994ApJ...424..627E, 1998ApJ...500..147S},
which corresponds to a J-band extinction of about 3 mag,
according to the standard interstellar extinction 
law \citep{1985ApJ...288..618R}. This is certainly an upper limit, 
since the high visual extinction of 10 mag 
is most likely confined to the
starburst ring, i.e. to radii smaller than 1 kpc, while the fitted bulge
model reflects mean properties of a region extending far beyond
1 kpc.
Applying these corrections results in estimates about 
the lower limits for the M/L (Table~\ref{tab:gasdust}).
\begin{table*}
\caption{Estimated lower limits for the M/Ls of bulge and
disk after applying corrections for gas mass and dust extinction. The table
lists the component (Col. 1), the estimated gas mass expressed as a percentage
of the total mass (Col. 2), the estimated mean J-band extinction (Col. 3), the
resulting lower limit for the corrected M/L (Col. 4), 
and the original (uncorrected) value of the M/L (Col. 5) for comparison.}
\label{tab:gasdust}
$$
\begin{array}{lcccc}
	\hline
	\noalign{\smallskip}
	\mathrm{Component} & \mathrm{Gas\ Mass} 
	& \mathrm{J-Band\ Extinction}   
	& \mathrm{(M/L)_{lower}}\ [M_{\sun}/L_{\sun}] 
	& \mathrm{(M/L)_{uncorrected}}\ [M_{\sun}/L_{\sun}] \\
	\noalign{\smallskip}
	\hline
	\noalign{\smallskip}
	\mathrm{disk} & \leq 7\% & \mathrm{\sim 0.5\ mag} & 
	0.4 & 0.7\pm 0.4 \\
	\mathrm{bulge} & \leq 20\% & \mathrm{\sim 3.0\ mag} & 
	0.02 & 0.4\pm 0.2 \\
	\noalign{\smallskip}
	\hline
\end{array}
$$
\end{table*}

\subsection{Stellar Populations}

Measured J-band M/Ls are very rarely found in literature but 
\citet{2000AJ....120.2884P} report bulge and disk M/Ls for a sample of 
74 spirals in the I-band. For comparison, the bulge and disk M/Ls of the
\object{I Zw 1} host have to be multiplied by a factor of about 2.3, 
according to
the host $\mathrm{I}-\mathrm{J}$ colour of roughly 0.9 mag 
\citep[][ Fig. 9]{1998ApJ...500..147S}. The M/Ls for the \object{I Zw 1}
host are similar to the majority of M/Ls found by 
\citet{2000AJ....120.2884P}, 
since most of their disk M/Ls  are in the range of 
$1.0\ M_{\sun} / L_{\sun}$ to
$3.0\ M_{\sun} / L_{\sun}$ and most of their bulge M/Ls are significantly lower
with values below $1.0\ M_{\sun} / L_{\sun}$.
 
As the M/Ls increase with increasing age of the underlying stellar population,
the derived M/Ls for the \object{I Zw 1} host 
can be used to estimate the mean stellar population in the disk and the bulge
component. 
The 
derived M/L of the disk of 
$\lesssim (0.7\pm 0.4)\ M_{\sun} / L_{\sun}$ is slightly sub-solar. 
It 
indicates a predominance of normal stars with a tendency towards a slightly 
increased fraction of younger stars. Such a tendency could be explained
by the enhanced star formation activity found for the north-western
spiral arm \citep{1998ApJ...500..147S}.
The M/L of the bulge of $\mathrm{\lesssim (0.4\pm 0.2)}\ M_{\sun} / L_{\sun}$ 
is much lower and
hints at a young stellar population with a significant fraction 
of hot stars and supergiants. Such an interpretation is supported by 
the finding of strong starburst activity
in the circumnuclear region of \object{I Zw 1} \citep{1998ApJ...500..147S, 
2001IAUS..205..340S}.

\section{Summary and Conclusions\label{sec:summary}}

As a nearby QSO, yet showing many properties of high-redshift
counterparts, and as a likely transition object in the 
original evolutionary scheme for AGN, \object{I Zw 1} is 
a very interesting candidate for a detailed case study of a QSO host.
This paper presents new results from J-band imaging 
of the \object{I Zw 1} host with ISAAC at the VLT
-- the first investigations in a series, turning \object{I Zw 1} into 
one of the best-studied QSO hosts.

The analysis shown here provides first-time results focussing on the 
structure of the underlying host galaxy of \object{I Zw 1}, since the  
host is
uncovered by a reliable subtraction of the QSO nucleus from the 
J-band images. By the structural decomposition 
of the surface brightness profile and the gas rotation curve
the host is characterized as a Freeman-type-I high surface brightness galaxy.
It is shown that the radial distribution of luminous matter is 
similar to that found for 
nearby AGN spiral hosts. 
Different solutions for the decomposition of the rotation curve are discussed
assuming swing amplification as the prevailing mechanism
for the amplification the two-armed spiral structure 
in the \object{I Zw 1} disk.
The resulting M/Ls indicate a tendency towards a younger stellar population 
in the disk and a predominance of young hot stars and supergiants in the
bulge. This interpretation aims at the same direction as previous findings
about enhanced star formation in the north-western spiral arm and strong 
starburst activity in the circumnuclear molecular ring of \object{I Zw 1}.
New evidence for a tidal interaction between \object{I Zw 1} and 
the nearby western companion galaxy is found in the sensitive high-resolution
ISAAC J-band image, which shows an elongation
of the companion as well as a tidal bridge and tail. 
Without a large-scale bar in the \object{I Zw 1} host, 
this interaction seems to be the major mechanism for triggering the inflow of
gas which is needed to fuel the circumnuclear starburst as well as the 
QSO. 
The anticipated merger process would be a further argument for \object{I Zw 1}
to be a transitional object in the original evolutionary scheme. 

\begin{acknowledgements}
We would like to thank the observer who carried out the VLT observations in 
service mode.
We also thank Prof. Dr B. Fuchs for helpful comments and suggestions.
J. Scharw\"achter is supported by a scholarship for doctoral students of the
``Studienstiftung des deutschen Volkes''.
\end{acknowledgements}

\bibliographystyle{aa}
\bibliography{h4270.bib}

\end{document}